\documentclass[10pt,twocolumn,english,aps,reprint,showpacs,longbibliography,footnotebib,superscriptaddress]{revtex4-1}
\usepackage[T1]{fontenc}
\usepackage[latin9]{inputenc}
\usepackage{amsmath}
\usepackage{graphicx}
\usepackage{amssymb}

\makeatletter
\@ifundefined{textcolor}{}
{%
 \definecolor{BLACK}{gray}{0}
 \definecolor{WHITE}{gray}{1}
 \definecolor{RED}{rgb}{1,0,0}
 \definecolor{GREEN}{rgb}{0,1,0}
 \definecolor{BLUE}{rgb}{0,0,1}
 \definecolor{CYAN}{cmyk}{1,0,0,0}
 \definecolor{MAGENTA}{cmyk}{0,1,0,0}
 \definecolor{YELLOW}{cmyk}{0,0,1,0}
 }

\usepackage{float}\usepackage{babel}

\makeatletter
\@ifundefined{textcolor}{}
{%
 \definecolor{BLACK}{gray}{0}
 \definecolor{WHITE}{gray}{1}
 \definecolor{RED}{rgb}{1,0,0}
 \definecolor{GREEN}{rgb}{0,1,0}
 \definecolor{BLUE}{rgb}{0,0,1}
 \definecolor{CYAN}{cmyk}{1,0,0,0}
 \definecolor{MAGENTA}{cmyk}{0,1,0,0}
 \definecolor{YELLOW}{cmyk}{0,0,1,0}
 }

\makeatother

\makeatother

\usepackage{babel}

\begin{document}

\title{Natural human mobility patterns and spatial spread of infectious
diseases}

\author{Vitaly Belik}
\email{e-mail:belik@mit.edu}

\affiliation{Max Planck Institute for Dynamics and Self-Organization, G{\"o}ttingen, Germany}
\affiliation{Massachusetts Institute of Technology, Department of Civil and Environmental Engineering, Cambridge, MA, USA}
\author{Theo Geisel}
\affiliation{Max Planck Institute for Dynamics and Self-Organization, G{\"o}ttingen, Germany}
\affiliation{Institute for Nonlinear Dynamics, Department of Physics, University of G{\"o}ttingen, G{\"o}ttingen, Germany}
\author{Dirk Brockmann}
\affiliation{Northwestern Institute on Complex Systems, Northwestern University, Evanston, IL, USA}

\affiliation{Department of Engineering Sciences and Applied Mathematics, Northwestern University, Evanston, IL, USA}

\date{\today}

\begin{abstract}
We investigate a model for spatial epidemics explicitly taking into
account bi-directional movements between base and destination locations
on individual mobility networks. We provide a systematic analysis
of generic dynamical features of the model on regular and complex
metapopulation network topologies and show that significant dynamical
differences exist to ordinary reaction-diffusion and effective force
of infection models. On a lattice we calculate an expression for the
velocity of the propagating epidemic front and find that in contrast
to the diffusive systems, our model predicts a saturation of the velocity
with increasing traveling rate. Furthermore, we show that a fully
stochastic system exhibits a novel threshold for attack ratio of an
outbreak absent in diffusion and force of infection models. These
insights not only capture natural features of human mobility relevant
for the geographical epidemic spread, they may serve as a starting
point for modeling important dynamical processes in human and animal
epidemiology, population ecology, biology and evolution. 
\end{abstract}
\maketitle
The geographic spread of emergent infectious diseases, epitomized
by the 2009 H1N1 outbreak and subsequent pandemic~\cite{Fraser:2009p5192},
the worldwide spread of SARS in 2003~\cite{HBGPnas,Colizza:2007p1229},
and recurrent outbreaks of influenza epidemics~\cite{Balcan:2009fy,Balcan2010,Balcan_pnas},
is determined by a combination of disease relevant human interactions
and mobility across multiple spatial scales~\cite{Rileyreview}.
While infectious contacts yield local outbreaks and proliferation
of a disease in single populations, multiscale human mobility is responsible
for spatial propagation~\cite{may}. Therefore, a prominent lineage
of mathematical models has evolved that is based on reaction-diffusion
dynamics~\cite{kpp,fisher} in which the combination of local exponential
growth and diffusive dispersal captures qualitative aspects of observed
dynamics. Typically these systems exhibit constant velocity epidemic
wavefronts. A related class of phenomenological models is based on
the concept of an effective force of infection across distance and
thus does not require explicit modelling of dispersal~\cite{Rushton1955,hagernaas}. 

\begin{figure}[t]
\begin{centering}
\includegraphics[width=1\columnwidth]{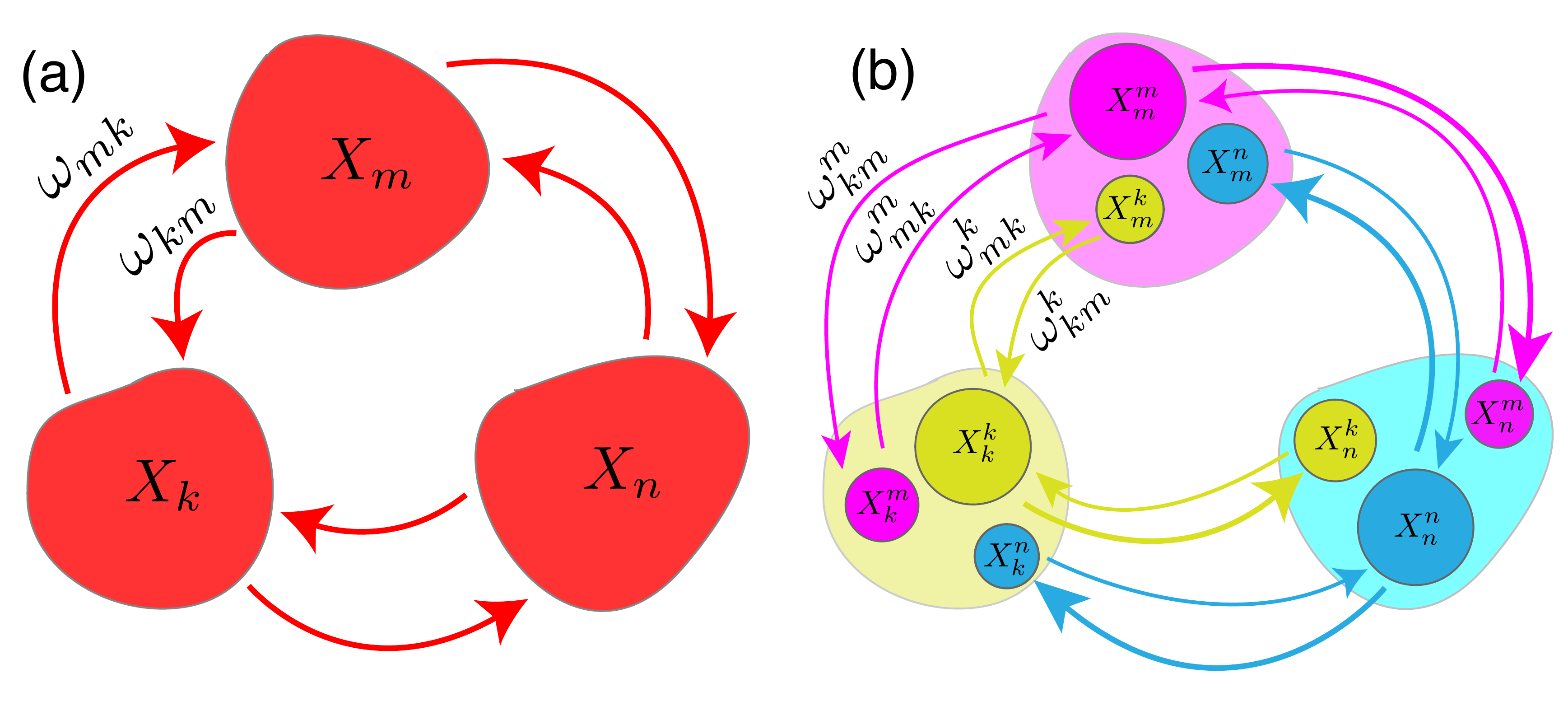} 
\par\end{centering}

\caption{(color online) Metapopulational mobility models. Patches and arrows
represent individual populations and travel. (a) Diffusive dispersal:
indistinguishable individuals travel randomly between different locations
governed by the set of transition rates $\omega_{nm}$. (b) Dispersal
capturing individual mobility patterns. Individuals with label $k$
possess travel rates $\omega_{mk}^{k}$ and $\omega_{km}^{k}$ at
which they travel from their base location $k$ to connected locations
$m$ and back.\label{fig. 1}}

\end{figure}

Although more sophisticated models~\cite{rvachev,HBGPnas,Vesp1,Balcan:2009fy}
have been developed to describe the dynamics of recent emergent epidemics
such as the H1N1 pandemic or SARS, taking into account long distance
travel and multiscale mobility networks~\cite{Vespignani:2009cq},
the majority of these models are still based on the interplay of local
reaction kinetics and diffusion processes on networks of metapopulations.
The key assumptions of diffusive transport (Fig.\ref{fig. 1} (a))
are that a) individuals behave identically, b) movements are stochastic,
c) spatial increments are local and as consequence individuals eventually
visit every location in the system. Although it is intuitively clear
that these assumptions are idealizations in conflict with everyday
experience, the difficulty is how to refine them when data on mobility
is lacking, is insufficient or incomplete. Fortunately, a series of
recent studies~\cite{DBNature,Gonzalez2008,Chaoming-Song-and-Zehui-Qu-and-Nicholas-Blumm-and-Albert-Laszlo-Barabasi-:2010la,Song:2010kx}
substantially advanced our knowledge on multiscale human mobility.
An important discovery that emerged from these studies, are individual
mobility networks, i.e. individuals typically only visit a limited
number of places frequently, predominantly performing commutes between
home and work locations and possibly a few other locations. Consequently
individuals or groups of individuals exhibit spatially constrained
movement patterns despite their potentially high mobility rate. It
has remained elusive how this novel and important empirical insight
on individual mobility networks can be reconciled with epidemiological
models, to which extent it may impact spatial disease dynamics, and
how it may alter spreading scenarios and predictions promoted by ordinary
reaction-diffusion models, in which mobile hosts can reach every location
in the system.

In this Letter we demonstrate how individual mobility networks can
be incorporated into a class of models for spatial disease dynamics,
and address the questions whether and how significantly the existence
of spatially constrained individual mobility networks impact on key
features of disease dynamics. To this end we investigate a model that
explicitly accounts for bidirectional mobility of individuals between
their unique base location (e.g. their home) and a small set of other
locations~(Fig.\ref{fig. 1}(b)). The entire population is therefore
represented by a set of overlapping individual mobility networks associated
with each base location, see e.g.~\cite{bidirect1,Keeling2002,Arino}.
We focus on the analysis of epidemics on regular lattices and complex
metapopulation networks and systematically compare the dynamics to
reaction-diffusion systems as well as force of infection models. In
conflict with reaction-diffusion models, in which front velocities
increase with travel rates unboundedly, we show that our model predicts
a saturation of wave front velocities, a direct consequence of the
rank of locations in individual mobility patterns. This suggests that
estimates for propagation speeds may have been considerably overestimated
in the past. Furthermore, we analyze a fully stochastic model to show
that both, in regular lattices as well as complex metapopulation networks
the global outbreak of a disease is determined by a novel type of
threshold for the attack ratio of the disease which depends on the
charactetistic time spent at distant locations. Finally, we show that
in the limit of low and high travel rates our model agrees with reaction-diffusion
models and the direct force of infection class of models, respectively.

Consider a system of $M$ populations labeled $m$ and assume that
in each an epidemic outbreak can be described by a compartmental SIR-model\begin{equation}
I_{m}+S_{m}\xrightarrow{\alpha}2I_{m},\quad I_{m}\xrightarrow{\beta}R_{m},\label{eq:hund}\end{equation}
where $S_{m},I_{m}$ and $R_{m}$ label and quantify susceptible,
infected and recovered individuals of population $m$. Infections
and recovery events occur at rates $\alpha$ and $\beta$, respectively,
with $\alpha>\beta$. The number of individuals in population $m$
is given by $N_{m}=S_{m}+I_{m}+R_{m}$, which though is conserved
only in statistical sense at equilibrium. This system of reactions
yields the mean-field description $\partial_{t}I=\alpha IS/N-\beta I$
and $\partial_{t}S=-\alpha IS/N$. A natural and plausible extension
to a system of $M$ coupled populations is diffusive dispersal among
those populations defined by a hopping rates $w_{nm}>0$ from population
$ $$m$ to $n$, which yields a metapopulation reaction-diffusion
system, i.e. (for infectives and susceptibles) \begin{eqnarray}
\partial_{t}I_{n} & = & \alpha S_{n}I_{n}/N_{n}^{s}-\beta I_{n}+\sum_{m}\left(w_{nm}I_{m}-w_{mn}I_{n}\right)\nonumber \\
\partial_{t}S_{n} & = & -\alpha S_{n}I_{n}/N_{n}^{s}+\sum_{m}\left(w_{nm}S_{m}-w_{mn}S_{n}\right)\label{eq:dieter}\end{eqnarray}
where $N_{n}^{s}$ is the number of individuals in population $n$
in diffusive equilibrium. Travel rates $\omega_{mn}$ are usually
estimate by gravity-like laws~\cite{BelikBrockmann}. $N_{n}^{s}$
is determined by detailed balance, i.e. $N_{n}^{s}/N_{m}^{s}=w_{nm}/w_{mn}$
and conservation of the number of individuals in the metapopulation
$\mathcal{N}=\sum_{m}N_{m}^{s}$ . These type of models have been
employed in numerous recent studies~\cite{HBGPnas,CollizaVespignani2007,Vesp1,stratAsia}.
The tight connection to spatially continuous reaction diffusion systems
is revealed for the special case of a linear grid of populations placed
at regular intervals $l$ at positions $x_{m}=ml$ and mobility between
neighboring populations, i.e. $w_{nm}=\omega\delta_{n-1,m}+\omega\delta_{n+1,m}$
which yields\begin{eqnarray}
\partial_{t}j & = & \alpha js-\beta j+D\partial_{x}^{2}j\label{eq:RD}\\
\partial_{t}s & = & -\alpha js+D\partial_{x}^{2}s\nonumber \end{eqnarray}
 in which $j(x,t)=I_{n}/N_{n}^{s}$, $s(x,t)=S_{n}/N_{n}^{s}$ and
$D=l^{2}w$. Eq.~(\ref{eq:RD}) is related to a classic form of a
Fisher equation which has been deployed in mathematical epidemiology~\cite{kpp,fisher}.
For sufficiently localized initial conditions this systems exhibits
travelling waves with speed\begin{equation}
c=2\sqrt{\alpha D(1-\beta/\alpha)}\sim\sqrt{\omega}.\label{eq:FKV}\end{equation}
 Note that this velocity increases monotonically with the mobility
rate $\omega$. %

In order to account for individual mobility patterns we propose the
following generalization: individuals possess two indices $n$ and
$k$, characterizing their current and their base location, respectively.
The dispersal dynamics is governed by a Markov process\begin{equation}
X_{n}^{k}\overset{\omega_{mn}^{k}}{\underset{\omega_{nm}^{k}}{\rightleftharpoons}}X_{m}^{k},\label{eq:horst}\end{equation}
 where $X$ stands for each infectious state $S$, $I$ and $R$.
Eqs.~(\ref{eq:horst}) imply that individuals with base location
$k$ possess a unique mobility rate $w_{nm}^{k}$ that determines
how they travel between locations $n$ and $m$. This yields a generalization
of~\eqref{eq:dieter}:\begin{eqnarray}
\partial_{t}I_{n}^{k} & = & \frac{\alpha}{N_{n}}S_{n}^{k}\sum_{m}I_{n}^{m}-\beta I_{n}^{k}+\sum_{m}\left(\omega_{nm}^{k}I_{m}^{k}-\omega_{mn}^{k}I_{n}^{k}\right)\nonumber \\
\partial_{t}S_{n}^{k} & =- & \frac{\alpha}{N_{n}}S_{n}^{k}\sum_{m}I_{n}^{m}+\sum_{m}\left(\omega_{nm}^{k}S_{m}^{k}-\omega_{mn}^{k}S_{n}^{k}\right)\label{eq:bi_mf}\end{eqnarray}
 where $I_{n}^{k}$ and $S_{n}^{k}$ are the number of infecteds and
susceptibles of type $k$ that are currently located in population
$n$. $N_{n}$ is the total number of individuals at $n$, i.e. $N_{n}=\sum_{k}(I_{n}^{k}+S_{n}^{k}+R_{n}^{k})$.
If $\omega_{nm}^{k}$ are $k$-independent, we recover the reaction-diffusion
case described above. In the following we consider the case of overlapping
star-shaped networks corresponding to commuting between base and destination
locations only. This implies that $\omega_{nm}^{k}=0$ if either $k\neq n$
or $k\neq m$. We further assume the constant return rate $w_{mk}^{k}=w^{-}$
for all $k$ and $m$. Realistically we have $\omega_{kn}^{n}/\omega^{-}\ll1$,
implying individuals belonging to $n$ remain at their base most of
the time.

If mobility rates are large compared to the rates associated with
the infection and recovery, i.e. $\omega_{mk}^{k},\omega^{-}\gg\alpha,\beta$,
detailed balance is fulfilled for infecteds and susceptibles separately
and the last term in Eq.~\eqref{eq:bi_mf} vanishes in which case
the above model is equivalent to the remote force of infection model,
see also~\cite{Rushton1955,Keeling2002}.

In general, it is difficult if not impossible to extract the dynamic
differences between the systems defined by Eqs.~\eqref{eq:dieter}
and~\eqref{eq:bi_mf} for an arbitrary metapopulation. Yet, it is
crucial to understand the key dynamic differences that emerge when
individual mobility patterns replace ordinary diffusion. We therefore
investigate the impact of individual mobility patterns in the instructive
one-dimensional lattice case. We assume that only infecteds can travel
and that recovery is absent (SI model), i.e. $\beta=0$ (relaxing
these two restrictions does not change the main results but clarifies
the analysis). We denote the number of infecteds with base location
$n$ that are located on site $n$ (i.e. their base) and sites $n\pm1$
by $I_{n}^{n}$ and $I_{n}^{\pm}$ , respectively. At rates $\omega^{+}$
and $\omega^{-}$ individuals leave and return to their base. In order
to obtain a spatially continuous description we approximate $I_{n\pm1}^{\pm}\rightarrow I^{\pm}(x\pm l)\approx I^{\pm}\pm l\nabla I^{\pm}+\frac{l^{2}}{2}\Delta I^{\pm}$
and introduce relative concentrations $u_{n}=I_{n}/N$, $v_{n}=(I_{n}^{+}+I_{n}^{-})/N$
leading to\begin{eqnarray}
\partial_{t}u & = & \alpha(1-u-v)(u+v+\mathcal{D}\Delta v)+\omega^{-}v-2\omega^{+}u\nonumber \\
\partial_{t}v & = & 2\omega^{+}u-\omega^{-}v\label{eq:uv}\end{eqnarray}
 where $\mathcal{D}=l^{2}/2$. The function $u(x,t)$ is the fraction
of individuals based and located at $x$ whereas $v(x,t)$ is the
fraction of individuals based but not located at $x$. Note that we
discarded here a third equation for $w=(I^{+}-I^{-})/N$ independent
from \eqref{eq:uv} with a solution vanishing at long times. This
does not change the main result \eqref{eq:MV}. The non-trivial steady
state is given by $u^{\star}=\omega^{-}/(2\omega^{+}+\omega^{-})$,
$v^{\star}=2\omega^{+}/(2\omega^{+}+\omega^{-})$ with $u^{\star}+v^{\star}=1$
as expected. %
\begin{figure}[t]
\begin{centering}
\includegraphics[width=0.8\columnwidth]{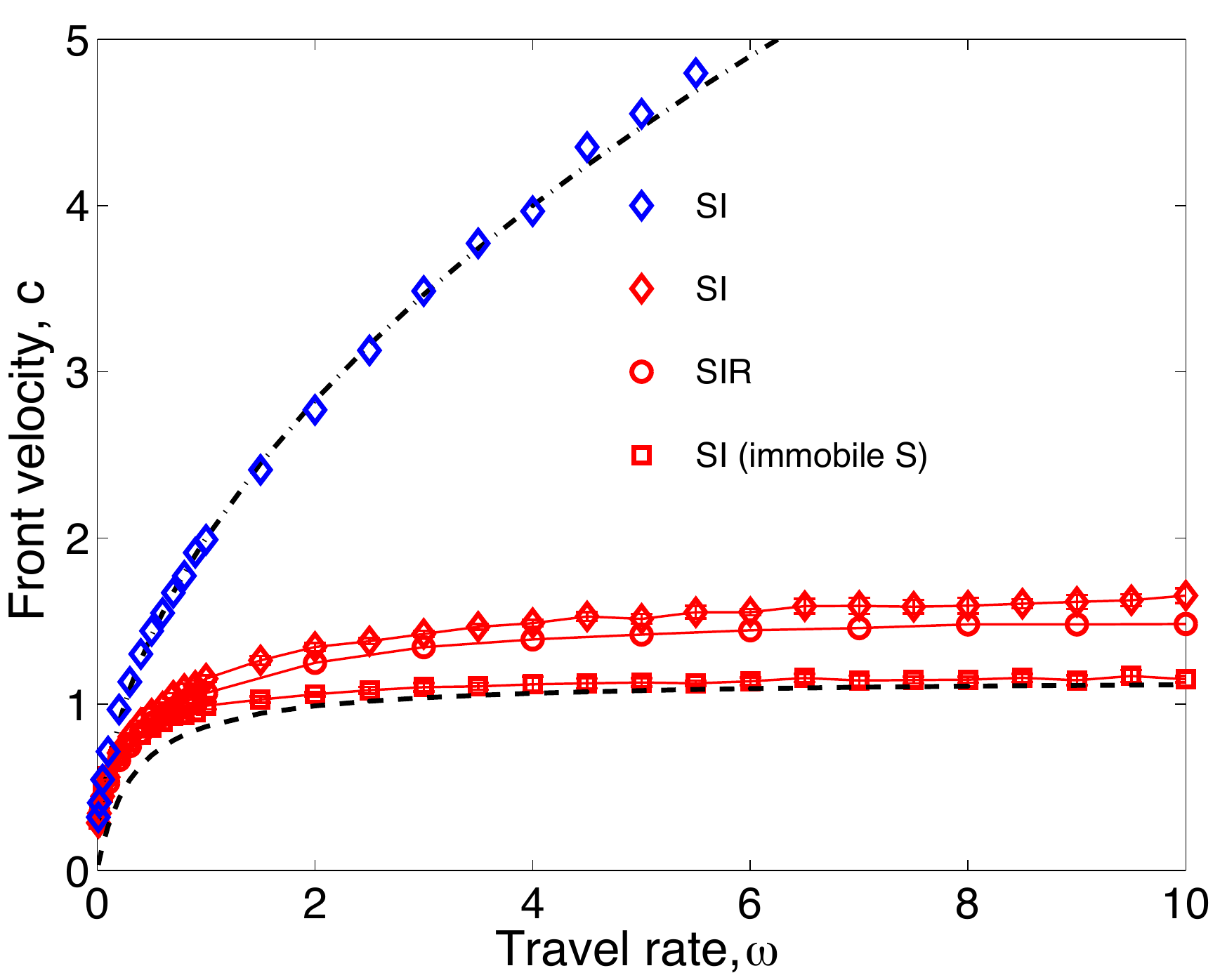}
\par\end{centering}

\caption{(color online) Front velocity $c(\omega)$ of the model defined by
Eq.~\eqref{eq:bi_mf} as a function of mobility rate $\omega$ in
comparison to ordinary reaction-diffusion dynamics (Eq.~\eqref{eq:dieter}).
For stochastic simulations we used a lattice with $10^2$ sites and $N=10^{4}$ individuals/site. Mean
velocities are indicated by red (bi-directional mobility) and blue
(reaction-diffusion) symbols. Analytical results, i.e. Eqs.~\eqref{eq:BDV}
and~\eqref{eq:FKV} are shown by dashed and dash-dotted lines respectively.
\label{fig 2}}

\end{figure}

One of the key questions is if the system (\ref{eq:uv}) exhibits
stable propagating waves as solutions and how their velocity depends
on system parameters. The ansatz $u(x,t)=\tilde{u}(x-ct)$ and likewise
for $v$, leads to a stable solution with a velocity given by\begin{equation}
c=\frac{2\alpha\omega^{+}\sqrt{\mathcal{D}\left(2+\omega^{-}/\omega^{+}\right)}}{\left(\alpha+\omega^{-}+2\omega^{+}\right)}.\label{eq:MV}\end{equation}
 Fixing $\omega^{-}$ this implies that $c\rightarrow0$ as $\omega^{+}\rightarrow0$.
On the other hand if $\omega^{-}$ is small, the system is entirely
determined by forward rate $\omega^{+}$. For a systematic comparison
to the ordinary reaction-diffusion system, we have to establish a
relation between $\omega^{\pm}$ and $\omega$ in Eq.~\eqref{eq:FKV}.
For the sake of simplicity we consider the symmetric case, i.e. $\omega^{+}=\omega^{-}=\omega$,
yielding \emph{\begin{equation}
c=\frac{2\sqrt{6\mathcal{D}}\alpha\omega}{\left(\alpha+3\omega\right)}.\label{eq:BDV}\end{equation}
 }Numerical simulations of a fully stochastic system nicely agree
with the analytical predictions (Fig.~\ref{fig 2}). The essential
feature of $c(\omega)$ is its saturation as $\omega\rightarrow\infty$
whereas $c\sim\sqrt{\omega}$ for reaction-diffusion systems. This
effect is a direct consequence of the spatial constraints imposed
by individual mobility patterns, i.e. increasing the mobility rate
$\omega$ yields a higher frequency of travel events but restricted
to the set of locations connected to the base location and thus is
universal and holds also for more complex metapopulation topologies~%
\footnote{We assume here that individuals return to their base location $k$
before travelling somewhere else. The initial model defined by Eqs.~\eqref{eq:horst}
allows to explore more general mobility patterns by relaxing this
assumption. However, as long as the individual rates $\omega_{nm}^{k}$
are dominated by the dynamics that incorporate a base location, propagation
speeds do not deviate significantly from the model we investigate
here.}. For high rates the deviation between ordinary reaction diffusion
and bi-directional mobility increases without bounds. This is an important
insight as the expression for wave front speeds for reaction-diffusion
have been used in the past to estimate wave front speeds from mobility
rates and vice versa. Our results indicate that if constrained mobility
patterns and bi-directional movement patterns are at work these estimates
must be treated with particular care. Note that in the limit $\omega\rightarrow\infty$
Eq.\emph{ }\eqref{eq:uv} reduces to the heuristic equation in Ref.~\cite{Post2}.

A key question in epidemiological contexts concerns conditions under
which an epidemic outbreak propagates or wanes. Outbreak dynamics
is usually triggered by crossing thresholds inherent in the system's
dynamics~\cite{Ball}. For example the basic reproduction number
$\mathcal{R}_{0}$ of a disease, i.e. the expected number of secondary
cases caused by a single infected individual in a susceptible population
represents such a threshold~\cite{may}. In single population SIR
dynamics introduced above $\mathcal{R}_{0}=\alpha/\beta$. If $\mathcal{R}_{0}>1$
an outbreak occurs, otherwise epidemic dies out. In the extended metapopulation
system with diffusive host movements a second threshold, known as
the global invasion threshold~\cite{CollizaVespignani2007}, is induced
by the flux of individuals between populations which must be sufficiently
high for an epidemic to spread throughout the metapopulation.

\begin{figure}[t]
 \includegraphics[width=0.9\columnwidth]{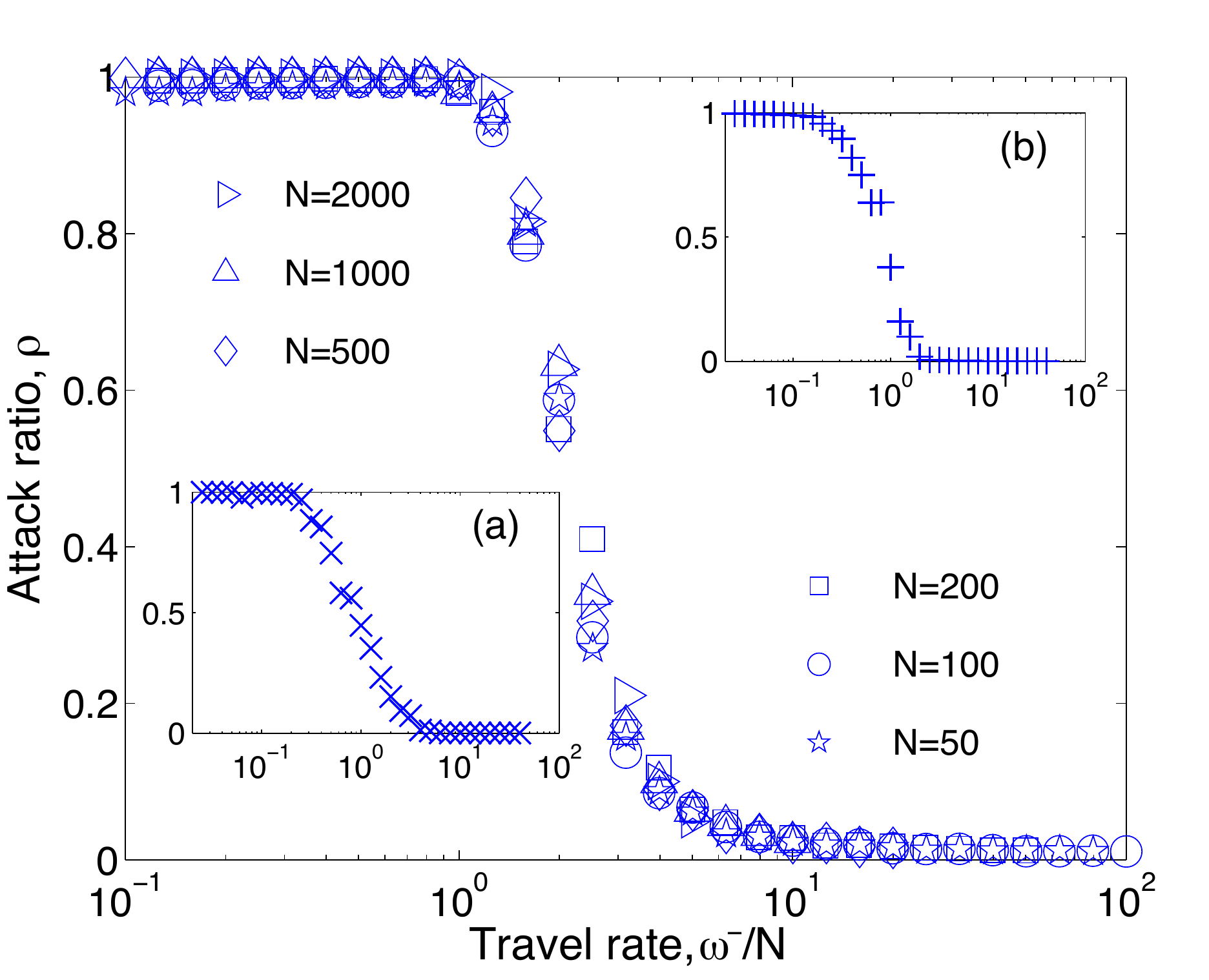} \caption{Results of stochastic simulations. Attack ratio $\rho(\omega^{-}/N)$
as a function of the return travel rate $\omega^{-}$ for SIR epidemics
on a lattice with $10^{2}$ sites with $N$ agents/site. Epidemic
parameters: $\alpha=1$, $\beta=0.1$. Insets: (a) --- $\rho(\omega^{-}/N)$
for a scale-free network ($\gamma=1.5$) of $10^{3}$ nodes populated
uniformly with $\langle N\rangle=250$/site, with $k_{\text{min}}=5$
and $k_{\text{max}}=50$; (b) --- $\rho(\omega^{-}/N)$ for an Erd\H{o}s-R{\'e}nyi
network with 500 nodes and $\bar{k}=10$. Results were averaged over
50 realisations. Interlocation flux rate was $\langle\omega\rangle=1$.\label{fig 3}}

\end{figure}

An interesting and important feature of the model we propose is the
existence of another invasion threshold which is determined by the
return rate $\omega^{-}$ or equivalently by the time an individual
spends outside the base location. This finding is illustrated in Fig.~\ref{fig 3}
which depicts the attack ratio $\rho$ as a function of $\omega^{-}$
for a bidirectional SIR epidemic on a lattice. The attack ratio $\rho$
is defined as the fraction of the overall population affected during
an epidemic. We fixed the entire interlocation flux at a value above
the global invasion threshold ensuring global outbreak in the reaction-diffusion
system. %

This result is universal for bidirectional system and is independent
of the particular topology. Insets (a) and (b) of Fig.~\ref{fig 3}
display simulation results for uncorrelated scale-free and Erd\H{o}s-R{\'e}nyi
networks. For low return rates the attack ratio is close to unity.
However, with growing values of the return rate, the attack ratio
decreases steeply and vanishes, reflecting the absence of the global
outbreak. The regime of high return rates corresponds to small dwelling
times on distant locations. Infected individuals do not have enough
time to transfer the disease to susceptibles in unaffected locations
before they return. Note that for a wide range of local population
sizes ($50-2,000$) the attach ratio $\rho$ collapses onto a universal
curve $\rho\sim\rho(\omega^{-}/N)$ which suggests that the basic
mechanism of the invasion threshold and its functional dependence
on return rate $\omega^{-}$ can be understood theoretically.

To estimate the observed invasion threshold on a one-dimensional lattice
for $\mathcal{R}_{0}\gtrapprox1$ we calculated the number of infecteds
$\lambda$ that originate from an affected location and can seed an
outbreak in an unaffected one~\cite{Ball}. These seeders are approximately
given by the total number of individuals entering the new location
per unit time, i.e. $N\omega$ multiplied by the typical time they
remain active in the destination location. This time is given by the
inverse rate $r$ of exiting the seeders class. Since two processes
(returning to the base location and recovery) can trigger this exit,
$r=\beta+\omega^{-}$. Therefore $\lambda\approx N\omega/(\beta+\omega^{-}).$
In the tree approximation the threshold condition for an epidemic
on a network with an average degree $\bar{k}$ is given by $\lambda(\bar{k}-1)(\mathcal{R}_{0}-1)>1$~\cite{CollizaVespignani2007}.
Thus, for a one-dimensional lattice the threshold condition reads
\begin{equation}
\frac{\omega^{-}}{\beta}\lesssim\left(2N(\mathcal{R}_{0}-1)\frac{\omega^{+}}{\beta}-1\right),\label{eq:Thresh}\end{equation}
where we kept the total interlocation flux $\omega=2\omega^{+}\omega^{-}/(2\omega^{+}+\omega^{-})$
constant (this relation and positivity of travel rates impose the
restrictions: $\omega^{-}>\omega$ and $\omega^{+}<\omega/2$) and
used $\omega^{+}\ll\omega^{-}$ . Moreover, as extensive simulations
show, the global invasion threshold in terms of the flux rate $\omega$
exists in bidirectional systems only for low return rates $\omega^{-}$.
With increasing return rate the disease fails to propagate globally,
even for constant interlocation flux $\omega$. This effect, observed
in lattice topologies and paradigmatic more realistic network topologies
is a fundamental consequence of birectional mobility patterns. This
effect is absent in reaction-diffusion systems fully determined by
the total interlocation flux. This property implies that invasion
of a front is limited only by return rates and not as is typically
assumed by the overall particle flux $\omega$. In the context of
disease dynamics this implies that more efficient containment strategies
could be potentially devised that do not target the overall mobility
but rather modifies mobility patterns in an asymmetric way.

Note that the transition takes place only in a system with a finite
number of agents per site, in the system with an infinite number of
individuals this effect would disappear as also confirmed in Fig.~\ref{fig 3}.
Indeed from Eq.~\eqref{eq:Thresh} for $\beta\ll\omega^{-}$ the
scaling $\rho\sim\rho\left(\frac{N\omega}{\omega^{-}}\right)$ follows
--- with increasing number of individuals per site $N$, the threshold
value of the return rate $\omega^{-}$ increases. 

Recently an unprecedented amount of detailed information on human
activity became available requiring revision of established models
and new more sophisticated ones. We investigate an epidemiological
model explicitly incorporating such important properties of human
mobility as individual mobility networks and frequent bidirectional
movements between home and destination locations. It manifests surprising
dynamical features such as the existence of bounded front velocity
and novel propagation threshold which are universal for any metapopulation
topology. As more detailed data continues to become available our
approach is a promising foundation for the further research.

D.B. and V.B. acknowledge support from the Volkswagen Foundation,
D.B. acknowledges EU-FP7 grant Epiwork. V.B and D.B would like to
thank W. Noyes, and H. Schl{\"a}mmer for invaluable comments. \bibliographystyle{apsrev4-1}
\bibliography{biblio}
 
\end{document}